\documentclass[aps,prl,twocolumn,showpacs,superscriptaddress,longbibliography]{revtex4-1}

\usepackage{graphicx}
\usepackage{dcolumn}
\usepackage{color}

\usepackage{eucal}
\usepackage{amssymb}

\usepackage{amsmath,amsthm}

\graphicspath{{../figs/}}

\begin{document}
\title{Capillary action in scalar active matter}

\author{Adam Wysocki}
\email{a.wysocki@lusi.uni-sb.de}
\author{Heiko Rieger}
\email{h.rieger@mx.uni-saarland.de}
\affiliation{Department of Theoretical Physics and Center for Biophysics, Saarland University, Saarbrücken, Germany}

\date{\today}
\begin{abstract}
  We study the capacity of active matter to rise in thin tubes against
  gravity and other related phenomena, like, wetting of vertical
  plates and spontaneous imbibition, where a wetting liquid is drawn into a porous medium.
  This capillary action or capillarity is well known in classical fluids and
  originates from attractive interactions between the liquid molecules and
  the container walls, and from the attraction of the liquid molecules among each other.
  We observe capillarity in a minimal
  model for scalar active matter with purely repulsive interactions, where
  an effective attraction emerges due to slowdown during collisions
  between active particles and between active particles and walls.
  Simulations indicate that the capillary rise in thin tubes is approximately proportional
  to the active sedimentation length $\lambda$ and that the wetting height of a vertical
  plate grows superlinear with $\lambda$. In a disordered porous medium the imbibition
  height scales as $\langle h\rangle\propto\lambda\phi_m$, where $\phi_m$ is its
  packing fraction.
\end{abstract}

\pacs{Bla}

\maketitle

{\it Introduction} -- Recently, active matter, which consists of particles
(motile microorganism or active colloids) that consume nutrients or fuel and convert it
into a persistent motion, has received a lot of attention due to its intrinsic
out-of-equilibrium character on the microscale \cite{marchetti2013rmp,elgeti2015rop}.
The simplest
representative of active matter are spherically symmetric, active Brownian particles
(ABPs) without alignment, however, with excluded volume interactions
\cite{fily2012prl,redner2013prl}.
Further representatives of the same class, also called scalar active matter,
are, for example, run-and-tumble particles \cite{tailleur2008prl,solon2015epjst}
and active lattice gas \cite{thompson2011jsp,sepulveda2017prl,whitelam2018jcp,kourbane2018prl}.
Such system, although far from equilibrium, are in some sense reminiscent of
a passive fluid with attractive interactions \cite{cates2015arcmp,farage2015pre},
since ABPs slow down during collisions and effectively attract each other.
As a result, ABPs undergo a
motility-induced phase separation (MIPS) into a coexisting dense
and dilute phase \cite{fily2012prl,redner2013prl,stenhammar2014sm,myself2014epl}.
The same mechanism is responsible for adhesion of ABPs
to repulsive walls, an effect called wall accumulation \cite{lee2013njp,elgeti2013epl,sepulveda2017prl}.

The investigation of capillary action, the ability of liquids to rise in thin
tubes against the action of gravity, has a long history and goes back to
Leonardo da Vinci \cite{deGennes2013}. It originates from the attractive
interactions between the liquid molecules and the container walls and
from the attraction of the liquid molecules among each other
(causing surface tension). The height of the liquid column in the tube is
governed by the balance between the gain in surface energy, which is proportional
to the difference of the solid–vapor and the solid–liquid surface tension, and 
the cost in gravitational energy \cite{deGennes2013,rowlinson2013}. The classical picture seems to
prohibit the appearance of capillarity in systems with purely repulsive interactions,
however, as mentioned above, scalar active matter display in some sense an equilibrium
behavior \cite{fodor2016prl} of an attractive fluid with an equation of state for the pressure
and equality of pressures in coexisting phases \cite{solon2015np,solon2015prl,das2019sr}.
Still, their interfacial properties are contradicting:
despite stable liquid-gas interfaces the surface tension
was found to be negative \cite{bialke2015prl,patch2018sm,solon2018njp}, which
is also true for solid-liquid interfaces \cite{zakine2019arxiv}.
All in all, the appearance of capillarity and related phenomena in active systems
is not trivial and was not investigated so far.

The behaviour of active particles in a homogeneous medium or near flat surfaces
is well-studied. However, the natural habitat for many motile bacteria are
porous media, such as soil, tissue, or biofilm. So far
the transport properties of interacting and non-interacting active particles in
heterogeneous media under both quiescent and flow conditions
\cite{chepizhko2013prl,reichhardt2014pre,pince2016nc,alonso2019prf} have been studied.
To the best of our knowledge, the classical experiment of spontaneous
imbibition of a liquid into a porous medium \cite{alava2004aip,gruener2012pnas}
has not been performed in the context of active fluids up to now. 

{\it Model} -- Several minimal off-lattice models of isotropic active
particles without alignment interaction have been proposed, e.g.,
active Brownian particles \cite{fily2012prl}, run-and-tumble particles \cite{solon2015epjst}
or active Ornstein-Uhlenbeck particles \cite{farage2015pre}.
By contrast, we use a lattice model of scalar active
matter, the so-called active lattice gas (ALG)
\cite{thompson2011jsp,sepulveda2017prl,whitelam2018jcp,kourbane2018prl},
which allows for simulation of large systems and, in addition, can be described
by exact hydrodynamic equations on macroscopic scales \cite{kourbane2018prl}.

We consider $N$ particles on a square lattice with
$N_x\times N_y=(nL_x)\times(nL_y)$
sites, where $1/n$ corresponds to grid spacing, as will be explained later.
Four types of particles $\sigma_{i_x,i_y}\in\{l,r,u,d\}$ corresponding to particles
in a left, right, up and down drifting state can occupy a lattice site
$(i_x,i_y)$, c.f. the sketch in Fig.~S1 in Supplemental Material \cite{SI}.
In a simple exclusion process the maximum occupation
number per site is $1$. We label immobile obstacle sites as $\sigma_{i_x,i_y}=w$
and use them in order to construct walls, tubes and porous matrices.
An empty site is indicated by $\sigma_{i_x,i_y}=0$. ALG evolves according to four
possible processes, see Supplemental Material \cite{SI} for the corresponding
microscopic equations for the evolution of the average occupation number:
1) Symmetric diffusion: For nearest neighbor pairs $(i_x,i_y)$ and $(j_x,j_y)$,
$\sigma_{i_x,i_y}\neq w$ and $\sigma_{j_x,j_y}\neq w$ are exchanged at rate $D$.
2) Self-propulsion: $\forall (i_x,i_y)$: a particle of type $\sigma_{i_x,i_y}=l$ in
$(i_x,i_y)$ jumps to the left to $(i_x-1,i_y)$ if $\sigma_{i_x-1,i_y}=0$
at rate $V_0/n$ and analogous for particles of type $r$, $u$ and $d$.
3) Sedimentation: $\forall (i_x,i_y)$: particles of all types in $(i_x,i_y)$
jumps downwards to $(i_x,i_y-1)$ if $\sigma_{i_x,i_y-1}=0$ at rate $V_g/n$.
4) Tumbling: Particle switch state at rate $\alpha/n^2$.

Such rules generate a persistent random walk
\cite{bertrand2018prl,whitelam2018jcp} and the model exhibits motility-induced
phase separation about a critical activity and density as a result of excluded
volume interactions \cite{kourbane2018prl,whitelam2018jcp}.
We simulate the dynamics of the ALG via a random-sequential update together
with a rejection-free continuous-time Monte Carlo algorithm
\cite{bortz1975jcp}.
In detail, the update sequence is a randomly permuted set of the particle
labels, a process is chosen with probability $W/R$ and the time increment
after each move is $1/(NR)$, where $W$ denotes the rate of the chosen process
and $R=4D+V_0/n+V_g/n+3\alpha/n^2$ is the total rate.

With the above choice of rates and after diffusive rescaling of space $x=i/n$
and time $\tau=t/n^2$ all processes contribute equally to the macroscopic
description of the system \cite{masi1985prl}. Hydrodynamic equations for
the densities of the four species $\{l,r,u,d\}$ can be derived
in case of large $n$ with diffusion constant $D$, tumbling rate $\alpha$,
propulsion $V_0$ and sedimentation velocity $V_g$, see Eq.~(17-20), 
in Supplemental Material \cite{SI}.

Important dimensionless numbers are the active $Pe_{a}=V_0/\sqrt{D\alpha}$
and gravitational P\'eclet number $Pe_{g}=V_g/\sqrt{D\alpha}$,
which compare either active swimming or gravitation-induced drift motion
to thermal diffusion. We use $l=\sqrt{D/\alpha}$ as a length scale.

Our basic setup consists of a ALG confined between horizontal walls at
$y=0$ and $y=L_y$, we apply periodic boundary conditions along $x$-direction
and gravity acts along the negative $y$-direction.
As a consequence, a dense phase covers the lower wall with a dilute phase
on top of it. The density profile of the dilute phase decays exponentially
as $\rho(y)\propto\exp{(-y/\lambda)}$, where $\lambda$ is the active
sedimentation length, which scales as
\begin{equation}
\lambda\propto\frac{V_0^2}{\alpha V_g}
\end{equation}
for large activity $Pe_a$ \cite{tailleur2009epl,enculescu2011prl,solon2015epjst,ginot2018njp},
as have been confirmed by our simulations.
In the absence of gravity active particles would accumulate symmetrically
at both walls and  in case of an ideal ALG at large $Pe_a$ the density profile,
for instance, at the lower wall $y=0$ would decay as
$\rho(y)\propto C_1\exp{(-y/\lambda_1)}+C_2\exp{(-y/\lambda_2)}$ with
$\lambda_1=D/V_0$ and $\lambda_2=\sqrt{D/2\alpha}$, which can be calculated
exactly using the corresponding hydrodynamic equations \cite{SI}.

We insert a capillary or a porous matrix into the dense
phase (or confine the system along $x$-direction by vertical walls),
let the system evolve until steady state is reached and adjust simultaneously
the number of particles in order to fix the height of the interface between
the dense and the dilute phase in the bulk region (far away from perturbations
due to vertical walls or porous matrix). We define the position of the interface
as the isodensity curve $\rho(x,y)=0.6$, where $\rho=\rho_l+\rho_r+\rho_u+\rho_d$
is the total density, and fix the height of the interface in the bulk region
in all simulations to $y_{bulk}/l=4$. In the following, we measure heights,
e.g., the height of the meniscus $\Delta h$ in case of capillary rise, relative
to $y_{bulk}$. We choose the size of the system, such that $L_x$ and $L_y$ are
much larger than any other length scale in system, like swimming persistence
length $V_0/\alpha$ or active sedimentation length $\lambda$. A bulk ALG (no walls)
does not phase separate below a critical value $Pe_a^c=8$ \cite{mangeat2019private}.

Because the results of the Monte Carlo simulations with a grid resolution of
$n=10$ match results with $n=20$ and are consistent with the numerical solutions,
obtained using finite element method \cite{hecht2012jnm}, of the hydrodynamic equations
\cite{SI} corresponding to the limit $n\rightarrow\infty$, we used a resolution
of $n=10$ in the following.

\begin{figure*}[ht]
\centering
\includegraphics[width=1.75\columnwidth]{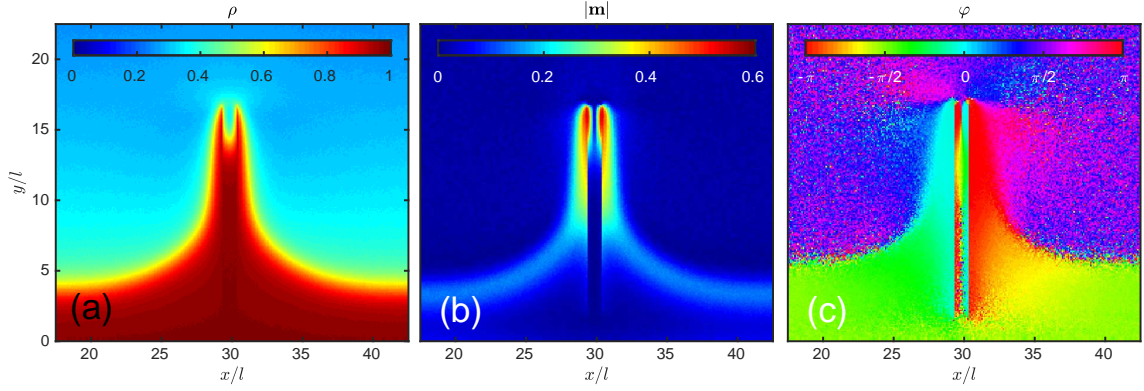}
\caption{\label{f:rho_capillary}(color online) Capillary rise of active lattice
  gas at active P\'eclet number $Pe_{a}=10$, gravitational P\'eclet number
  $Pe_{g}=0.2$, capillary width $\delta x/l=1$ and capillary height
  $\delta y/l=12.5$. The total size of the system is $L_x/l=60$ and $L_y/l=120$,
  which corresponds to $N_x\times N_y=1200\times2400$ lattice sites
  used in the Monte Carlo simulation. (a) Total density $\rho(x,y)\in [0,1]$.
  (b) Absolute value $|\mathbf{m}|(x,y)$ of the normalized polarization field
  $\mathbf{m}$, see Eq.~\ref{eq:polarization}.
  (c) Phase $\varphi(x,y)\in[-\pi,\pi]$ of
  $\mathbf{m}=|\mathbf{m}|\cdot(\cos{\varphi},\sin{\varphi})$.
  A phase $\phi=\{0,\pi/2,\{\pi,-\pi\},-\pi/2\}$ corresponds to a polarization
  along $\{\hat{\mathbf{x}},\hat{\mathbf{y}},-\hat{\mathbf{x}},-\hat{\mathbf{y}}\}$
  or $\{right,up,left,down\}$, respectively.}
\end{figure*}

{\it Capillary rise} -- In order to study capillarity in active systems
we use our basic setup and insert a tube into the dense phase. A typical
result for a thin tube and at sufficiently high $Pe_{a}$ and small $Pe_{g}$
is shown in Fig.~\ref{f:rho_capillary} with the total density
$\rho(x,y)\in[0,1]$ in Fig.~\ref{f:rho_capillary}(a), the absolute value
of the normalized polarization field
\begin{equation}
  \mathbf{m}=\left(\begin{matrix}m_x\\m_y\end{matrix}\right)
    =\frac{1}{\rho}\left(\begin{matrix}\rho_r-\rho_l\\\rho_u-\rho_d\end{matrix} \right)
      \label{eq:polarization}
\end{equation}
in Fig.~\ref{f:rho_capillary}(b) and the phase $\varphi(x,y)\in[-\pi,\pi]$ of
$\mathbf{m}=|\mathbf{m}|\cdot(\cos{\varphi},\sin{\varphi})$ in
Fig.~\ref{f:rho_capillary}(c), which indicates the direction of the polarization
$\mathbf{m}$. The ALG wets the walls of the tube accompanied by a strong polarization
in the direction antiparallel to the surface normal, the dense phase fills the
tube about the level of the bulk and a concave meniscus develops within the tube.
Accumulation at boundaries \cite{elgeti2013epl,sepulveda2017prl,ostapenko2018prl}
and capillary condensation in slit pores \cite{ni2015prl,wittmann2016epl}
are a well-known effect in scalar active matter, what is new, is the rise of the
dense phase inside the tube against gravity.

\begin{figure}[ht]
\centering
\includegraphics[width=\columnwidth]{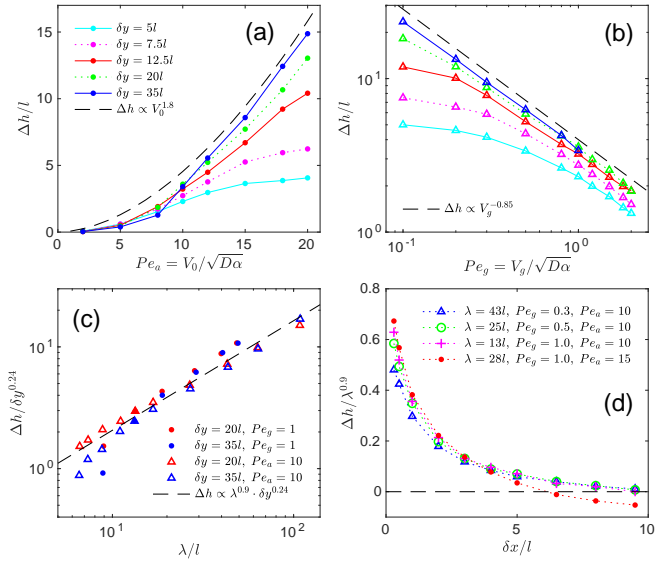}
\caption{\label{f:dh_lambda_dx}(color online) (a-c) The height of the meniscus
  $\Delta h$ for different tube heights $\delta y$ and at a fixed capillary width
  $\delta x/l=1$. (a) $\Delta h$ versus activity $Pe_{a}$ at fixed $Pe_{g}=1$ and
  (b) $\Delta h$ versus gravity $Pe_{g}$ at fixed $Pe_{a}=10$.
  (c) A master curve merging data from (a) and (b). $\Delta h$ is plotted
  as function of active gravitational length $\lambda$ and
  $\Delta h$ is scaled by $\delta y^{0.24}$. For every parameter pair $(Pe_{a},Pe_{g})$
  we estimate the corresponding $\lambda$ from a fit of the density profile $\rho(y)$
  in the bulk (far away from the tube) to $\rho(y)\propto\exp{(-y/\lambda)}$.
  Note that, at finite activity $\lambda$ is larger than the gravitational length
  in equilibrium $\lambda_{eq}=D/V_g$. (d) $\Delta h$ as function of the capillary
  width $\delta x$ for different parameter pairs $(Pe_{a},Pe_{g})$ and at fixed tube
  height $\delta y/l=17.5$. The height of the meniscus $\Delta h$ is scaled by
  $\lambda^{0.9}$.}
\end{figure}

Fig.~\ref{f:dh_lambda_dx}(a,b) illustrates the dependence of the height of the meniscus
$\Delta h$ on activity $Pe_{a}$ and gravity $Pe_{g}$ at fixed capillary width
$\delta x/l=1$. As intuitively expected, $\Delta h$ increases with activity and
decreases with gravity. Moreover, $\Delta h$ increases strongly with the tube height
$\delta y$ till some critical value and saturates above it. In the latter regime we observe
an approximate scaling $\Delta h\propto V_0^{1.8}$ and $\Delta h\propto V_g^{-0.85}$.
For every parameter pair $(Pe_{a},Pe_{g})$ we estimate the corresponding active gravitational
length $\lambda$ from a fit of the density profile $\rho(y)$ in the bulk
(far away from the tube) to $\rho(y)\propto\exp{(-y/\lambda)}$. Using this,
we obtain a master curve from data shown in Fig.~\ref{f:dh_lambda_dx}(a,b)
by plotting $\Delta h$ as a function of $\lambda$ and rescaling $\Delta h$ for
different tube heights by $\delta y^{0.24}$. As a result, we get a simple scaling relation
\begin{equation}
\Delta h\propto \lambda^{0.9},
\end{equation}
as indicated in Fig.~\ref{f:dh_lambda_dx}(c).

The height of the meniscus $\Delta h$ decreases strongly with increasing capillary
width $\delta x$, see Fig.~\ref{f:dh_lambda_dx}(d), and the curves for different parameter
pairs $(Pe_{a},Pe_{g})$ collapse after rescaling of $\Delta h$ with $\lambda^{0.9}$,
which is consistent with the master curve $\Delta h\propto\lambda^{0.9}$ in
Fig.~\ref{f:dh_lambda_dx}(c). For large $Pe_{a}$ and $\delta x$ the data collapse is not perfect,
where $\Delta h$ even becomes negative.

\begin{figure}[h]
\centering
\includegraphics[width=1\columnwidth]{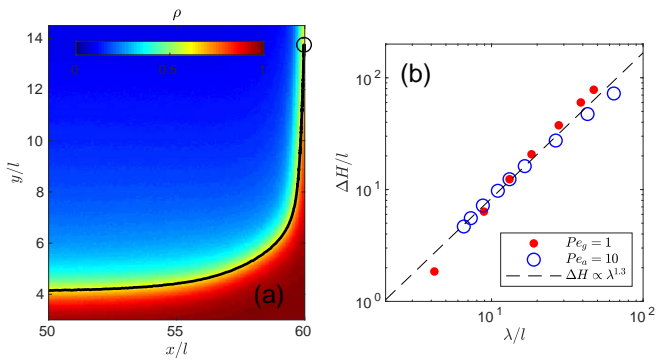}
\caption{\label{f:dH_lg}(color online) Wetting of a vertical plate.
  (a) Total density $\rho(x,y)\in [0,1]$ for $Pe_{a}=10$ and $Pe_{g}=1$. An cutout of
  a larger system ($L_x/l=60$ and $L_y/l=120$) is shown. Black line indicates the
  isodensity at $\rho=0.6$, which is our definition of the interface between
  the dense and the dilute phase. The height of the wetting layer $\Delta H$ is
  marked by a circle.
  (b) Height of the wetting layer $\Delta H$ as a function of the active gravitational
  length $\lambda$. Two data sets are shown: $2\le Pe_{a}\le 20$ at fixed $Pe_{g}=1$
  and $0.1\le Pe_{g}\le 2$ at fixed $Pe_{a}=10$.}
\end{figure}

{\it Wetting of a vertical plate} -- As mentioned above, the dense phase of the ALG not only
rises inside a capillary but also wets the outside walls of the capillary. In order to
study the latter effect separately we consider a setup of a rectangular container
with horizontal walls at $y=0$ and $y=L_y$ and vertical walls at $x=0$ and $x=L_x$.
Again we adjust the number of particles in order to fix the height of the interface
far away from walls ($x=L_x/2$). A typical total density field $\rho(x,y)\in [0,1]$
together with the isoline at $\rho=0.6$ (definition of the interface) is
shown in Fig.~\ref{f:dH_lg}(a), where a pronounced wetting of the vertical wall is
visible. We estimate the height of the wetting layer $\Delta H$ for different parameter
pairs $(Pe_{a},Pe_{g})$ together with the corresponding active gravitational lengths
$\lambda$ and obtaine a master curve, which suggests a superlinear growth
\begin{equation}
\Delta H\propto \lambda^{1.3}
\end{equation}
of the wetting height. Wetting can be understood as follows, activity leads to an
accumulation of particles at the container walls and gravity forces particles to sediment
to the bottom, however, excluded volume effect prevent a collapse of the wetting layer.
A zero-order approximation for the interface profile in a capillary tube of a width
$\delta x$ ia a superposition of wetting profiles of two independent walls at
distance $\delta x$.

\begin{figure}[h]
\centering
\includegraphics[width=.75\columnwidth]{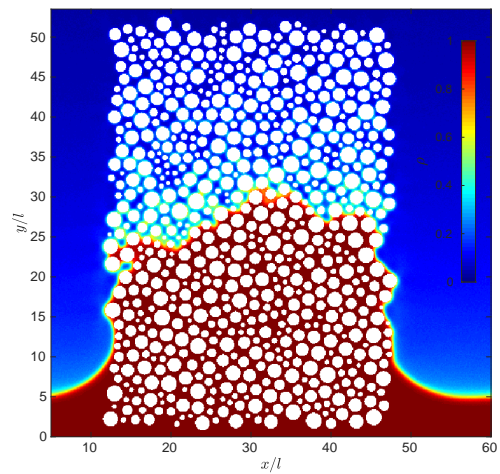}
\caption{\label{f:porous_snapshot}(color online)
  Imbibition of a porous media by the active lattice gas. The porous media
  consists of non-overlapping hard discs with uniformly distributed diameters
  in the range $\sigma/l\in[0.5,2]$; the minimum gap size is $0.325$ in units of $l$.
  The total density $\rho(x,y)\in [0,1]$ is shown for $Pe_{a}=15$, $Pe_{g}=1$
  and a system size $L_x/l=60$  and $L_y/l=100$. The porosity (or void fraction)
  is $1-\phi_m=0.2$, where $\phi_m$ is the packing fraction of the porous matrix.}
\end{figure}

{\it Imbibition of a porous matrix} -- Motivated by above results we probe how
the ALG penetrates an porous medium. We construct a porous matrix
from randomly placed non-overlapping hard discs with uniformly distributed
diameters in the range $\sigma/l\in[0.5,2]$ and a minimum gap size of $0.325l$.
As in the previous simulation setups, we adjust the number of particles in order
to fix the interface far away from the matrix. An example of a spontaneous imbibition
is shown in Fig.~\ref{f:porous_snapshot} for porosity (or void fraction)
of $1-\phi_m=0.2$ corresponding
to a packing fraction of the porous matrix of $\phi_m=0.8$. After an initial rise the
invasion front reaches a final height and width, and the interface fluctuates slightly
in the steady state.

\begin{figure}[h]
\centering
\includegraphics[width=.8\columnwidth]{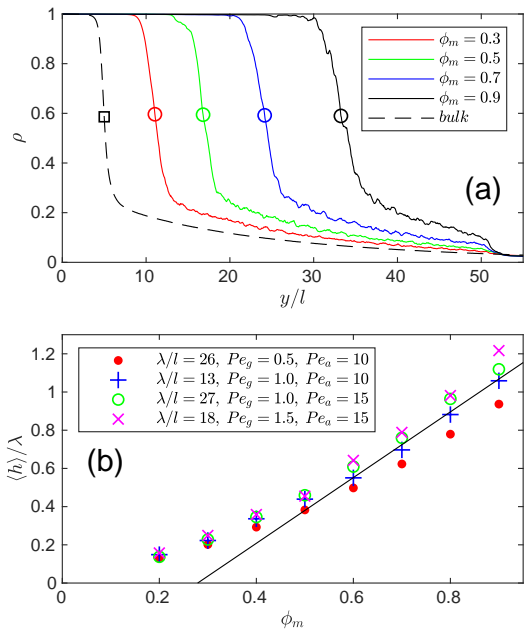}
\caption{\label{f:porous_profiles}(color online) (a) Density profiles $\rho(y)$
  of the active lattice gas within the porous matrix
  (the horizontal average was performed over the void region only)
  for different porosities $1-\phi_m$ (solid lines) and the corresponding $\rho(y)$
  in the bulk region (dashed line) are shown. The position of the interface is defined
  as the height where $\rho=0.6$ and is indicated by circles for the porous matrix
  and by a square in the bulk. The parameters are $(Pe_{a},Pe_{g})=(15,1)$.
  (b) Mean interface height $\langle h\rangle$ scaled by the active gravitational
  length $\lambda$ as a function of the packing fraction of the porous matrix
  $\phi_m$ for different parameter pairs $(Pe_{a},Pe_{g})$.}
\end{figure}

An obvious question is: What is the effect of the porosity on the imbibition?
In Fig.~\ref{f:porous_profiles}(a) we present the average density profiles
$\rho(y)$ of the active lattice gas within the porous matrix for different porosities
$1-\phi_m$ at $(Pe_{a},Pe_{g})=(15,1)$ and indicate the position of the interface by circles;
the horizontal average was performed over the void region only and over 10
independent realizations of the matrix. As expected, the smaller the porosity
the higher the rise of the dense phase within the porous matrix.
Interestingly, the density of the dilute phase above the invasion front
within the matrix is significantly larger then the bulk density at the same height,
see dashed line in Fig.~\ref{f:porous_profiles}(a).

We perform similar simulations for different $(Pe_{a},Pe_{g})$ and estimate
the mean interface height $\langle h\rangle$ and the corresponding
active gravitational length $\lambda$. A scaling of $\langle h\rangle$ with
$\lambda$ leads to a reasonable data collapse and indicates a simple growth relation
\begin{equation}
\langle h\rangle\propto\lambda\phi_m
\end{equation}
of the interface height, see Fig.~\ref{f:porous_profiles}(b).
A rough comparison between imbibition and capillary rise can be done by
plotting together the mean interface height $\langle h\rangle$ versus mean gaps size
of the porous matrix, obtained with Delaunay triangulation, and the height of the meniscus
$\Delta h$ versus capillary width $\delta x$, see Fig.~\ref{f:dh_lambda_dx}(d).
Although the shape of the curves is similar, the capacity of ALG to rise in porous media
against the action of gravity is twice as strong as in thin tubes.

{\it Conclusions and Outlook} -- We have show that, scalar active matter with purely
repulsive interactions can rise in thin tubes or invade porous matrix against
gravity. As a proof of concept we have used one of the simplest active matter models
and there are several obvious extensions and future directions in the study of active
capillarity and related phenomena using more realistic and elaborated models.
It is obvious to consider next off-lattice models, such as active Brownian particles (ABPs),
in order to test the universality of the scaling relations; preliminary simulations
clearly demonstrate capillary action in ABPs and confirm the robustness of the effect
in active matter systems. Particle-particle and particle-wall hydrodynamic
interactions \cite{drescher2011pnas,kantsler2013pnas}
or, in case of anisotropic active particles, alignment interactions can have a
dramatic effect on the behaviour near boundaries as compared to scalar active matter
and should be taken into account in future models.
It would be very interesting to investigate
the modification of classical capillary rise \cite{dimitrov2007prl} by activity
using active colloids with attraction \cite{redner2013pre},
where it is predicted that wall attraction may cause capillary drying \cite{wittmann2016epl}.
Also the study of capillary action and wetting in active multiphase system, such as
suspensions of motile bacteria \cite{lauga2016arfm}, algae \cite{goldstein2015arfm}
or synthetic self-propelled particles \cite{palacci2013science,buttinoni2013prl},
is very promising and could be a possible experimental test for our predictions.

\acknowledgements
This work was financially supported by the German ResearchFoundation (DFG)
within the Collaborative Research Center SFB 1027. We thank Matthieu Mangeat
for introduction into FreeFEM, Reza Shaebani for careful proofreading,
Sergej Rjasanow and Steffen Wei{\ss}er for tips about numerical methods.

\bibliography{refs}

\end{document}